\documentclass[preprint,aps,prd,groupedaddress,showpacs,floatfix,%
nofootinbib]{revtex4}

\catcode`\à=\active \defà{\`a} \catcode`\À=\active \defÀ{\`A}
\catcode`\á=\active \defá{\'a} \catcode`\Á=\active \defÁ{\'A}
\catcode`\ä=\active \defä{\"a} \catcode`\Ä=\active \defÄ{\"A}
\catcode`\â=\active \defâ{\^a} \catcode`\Â=\active \defÂ{\^A}
\catcode`\å=\active \defå{{\aa}} \catcode`\Å=\active \defÅ{{\AA}}
\catcode`\ç=\active \defç{\c{c}} \catcode`\Ç=\active \defÇ{\c{C}}
\catcode`\è=\active \defè{\`e} \catcode`\È=\active \defÈ{\`E}
\catcode`\é=\active \defé{\'e} \catcode`\É=\active \defÉ{\'E}
\catcode`\ë=\active \defë{\"e} \catcode`\Ë=\active \defË{\"E}
\catcode`\ê=\active \defê{\^e} \catcode`\Ê=\active \defÊ{\^E}
\catcode`\ì=\active \defì{\`{\i}} \catcode`\Ì=\active \defÌ{\`{\I}}
\catcode`\í=\active \defí{\'{\i}} \catcode`\Í=\active \defÍ{\'{\I}}
\catcode`\ï=\active \defï{\"{\i}} \catcode`\Ï=\active \defÏ{\"{\I}}
\catcode`\î=\active \defî{\^{\i}} \catcode`\Î=\active \defÎ{\^{\I}}
\catcode`\ò=\active \defò{\`o} \catcode`\Ò=\active \defÒ{\`O}
\catcode`\ó=\active \defó{\'o} \catcode`\Ó=\active \defÓ{\'O}
\catcode`\ö=\active \defö{\"o} \catcode`\Ö=\active \defÖ{\"O}
\catcode`\ô=\active \defô{\^o} \catcode`\Ô=\active \defÔ{\^O}
\catcode`\ù=\active \defù{\`u} \catcode`\Ù=\active \defÙ{\`U}
\catcode`\ú=\active \defú{\'u} \catcode`\Ú=\active \defÚ{\'U}
\catcode`\ü=\active \defü{\"u} \catcode`\Ü=\active \defÜ{\"U}
\catcode`\û=\active \defû{\^u} \catcode`\Û=\active \defÛ{\^U}
\catcode`\ý=\active \defý{\'y} \catcode`\Ý=\active \defÝ{\'Y}
\catcode`\ÿ=\active \defÿ{\"y} \catcode`\˜=\active \def˜{\"Y}
\catcode`\½=\active \def½{!`} \catcode`\¾=\active \def¾{?`}
\catcode`\ß=\active \defß{{\ss}}
\newcommand{\tmop}[1]{\ensuremath{\operatorname{#1}}}

\newcommand{\mathd}{\mathrm{d}}
\newcommand{\mathi}{\mathrm{i}}
\newcommand{\bignone}{}

\newcommand{\mathe}{\mathrm{e}}

\newcommand{\tmfloatcontents}{}
\newlength{\tmfloatwidth}
\newcommand{\tmfloat}[5]{
  \renewcommand{\tmfloatcontents}{#4}
  \setlength{\tmfloatwidth}{\widthof{\tmfloatcontents}+1in}
  \ifthenelse{\equal{#2}{small}}
    {\ifthenelse{\lengthtest{\tmfloatwidth > \linewidth}}
      {\setlength{\tmfloatwidth}{\linewidth}}{}}
    {\setlength{\tmfloatwidth}{\linewidth}}  \begin{minipage}[#1]{\tmfloatwidth}
    \begin{center}
      \tmfloatcontents
      \captionof{#3}{#5}
    \end{center}
  \end{minipage}}

\def\MeV{{\rm MeV}}
\usepackage{amsmath,amsfonts,bm}
\usepackage{amssymb}
\usepackage{dcolumn}

\usepackage{graphicx}
\usepackage{epsfig}
\usepackage{hyperref}
\usepackage{graphicx}
\usepackage{dcolumn}

\usepackage{longtable}
\begin{document}
\title{$\gamma^{*}\rho^0\to\pi^0$ Transition Form Factor in Extended AdS/QCD models}
\author{Fen
Zuo$^{a}$\footnote{Email: zuofen@itp.ac.cn},~Yu
Jia$^{b}$\footnote{Email: jiay@ihep.ac.cn}, and Tao
Huang$^{b}$\footnote{Email: huangtao@ihep.ac.cn}} \affiliation{
$^a$Key Laboratory of Frontiers in Theoretical Physics,~Institute of
Theoretical Physics,~Chinese Academy of Sciences,~Beijing 100190,
China\\
$^b$Institute of High Energy Physics and Theoretical Physics Center
for Science Facilities, Chinese Academy of Sciences,~Beijing 100049,
China}

\begin{abstract}
 The $\gamma^{*}\rho^0\to\pi^0$ transition form factor is extracted from
recent result for the $\gamma^* \gamma^* \pi^0$ form factor obtained
in the extended hard-wall AdS/QCD model with a Chern-Simons term. In
the large momentum region, the form factor exhibits a $1/Q^4$
behavior, in accordance with the perturbative QCD analysis, and also
with the Light-Cone Sum Rule (LCSR) result if the pion wave function
exhibits the same endpoint behavior as the asymptotic one. The
appearance of this power behavior from the AdS side and the LCSR
approach seem to be rather similar: both of them come from the
{``}soft" contributions. Comparing the expressions for the form
factor in both sides, one can obtain the duality relation $z\propto
\sqrt{u(1-u)}$, which is compatible with one of the most important
relations of the Light-Front holography advocated by Brodsky and de
Teramond. In the moderate $Q^2$ region, the comparison of the
numerical results from both approaches also supports a
asymptotic-like pion wave function, in accordance with previous
studies for the $\gamma^* \gamma^* \pi^0$ form factor. The form
factor at zero momentum transfer gives the $\gamma^{*}\rho^0\pi^0$
coupling constant, from which one can determine the partial width
for the $\rho^0(\omega)\to \pi^0 \gamma$ decay. We also calculate
the form factor in the time-like region, and study the corresponding
Dalitz decays $\rho^0(\omega)\to~\pi^0 e^+e^-,~\pi^0\mu^+\mu^-$.
Although all these results are obtained in the chiral limit,
numerical calculations with finite quark masses show that the
corrections are extremely small. Some of these calculations are
repeated in the Hirn-Sanz model and similar results are obtained.

\end{abstract}

 \maketitle

\section{Introduction}
In recent years, the phenomenological bottom-up approach to
describing strong interaction based on the AdS/CFT
correspondence~\cite{Maldacena1998,Witten1998a,Polyakov1998}, now
known as AdS/QCD, has offered much insight into various low-energy
aspects of QCD. In the simplest setup, various hadron states are
considered to be dual to different string modes propagating in a
slice of 5D AdS
space~\cite{Polchinski:2001tt,BoschiFilho:2002ta,deTeramond:2005su}.
High-energy scattering of glueballs naturally exhibits QCD-like
power behavior due to the warped geometry of the dual
theory~\cite{Polchinski:2001tt}. Spectra of low-lying hadron states
are well reproduced~\cite{deTeramond:2005su,Brodsky:2006uqa}. Chiral
symmetry and its spontaneous breaking are also well
implemented~\cite{Erlich:2005qh,DaRold:2005zs,Hirn:2005nr}. Up to
now there have been extensive studies on various dynamical
quantities such as decay constants, coupling constants and form
factors, e.g.,
\cite{Hong:2004sa,Grigoryan:2007vg,Brodsky:2007hb,Kwee:2007dd,Grigoryan:2007wn,Forkel:2007ru,Abidin:2009aj}.
Furthermore, a novel relation between the string modes and the
Light-Cone wave functions of the mesons was found in
Ref.~\cite{Brodsky:2006uqa}, from which the so-called Light-Front
holography was established.

To reproduce the Wess-Zumino-Witten term in the chiral Lagrangian, a
Chern-Simons~(CS) term must be
added~\cite{Sakai2005a,Panico:2007qd}. The CS term naturally
introduces baryon density~\cite{Domokos:2007kt}, since baryons are
related to the instantons in the 5D model.
The effect of this term to the baryon properties were later studied
in Ref.~\cite{Pomarol:2008aa}. Furthermore, with the CS term turned
on, the anomalous form factor of the pion coupling to two virtual
photon can be well reproduced~\cite{Grigoryan:2008up}.
Interestingly, the predictions for the form factor in the limit of
large photon virtualities coincide with those of perturbative
QCD~(pQCD) calculated using the asymptotic form of the pion
distribution amplitude. In this paper we attempt to extend this
calculation to the form factor of $\gamma^* \rho^0 \to \pi^0$
transition
, and then compare the results with those of the traditional
approaches.

In pQCD, the asymptotic behavior of the $\gamma^* \rho \pi$ form
factor has been predicted to be $1/Q^4$~\cite{Chernyak:1983ej}.
A simple expression for this form factor in large and moderate
momentum region can be obtained in the Light-Cone Sum Rules~(LCSR)
approach~\cite{Braun:1994ij,Khodjamirian:1997tk}, which gives the
same asymptotic behavior if the pion distribution amplitude is
asymptotic-like at the endpoint. However, the dominant contribution
is quite different from the pQCD analysis. At zero momentum transfer
the form factor defines the $\gamma \rho \pi$ coupling which
determines the width of the radiative decay $ \rho \to \pi \gamma$.
This coupling was extracted from the traditional three-point QCD sum
rule~\cite{Gokalp:2001sr}, and also from QCD sum rules in the
presence of external field~\cite{Zhu:1998bm}. In this paper we will
try to give a unified description of the form factor in the whole
region. We will mainly focus on the results in the standard
hard-wall model \cite{Erlich:2005qh,DaRold:2005zs}, and repeat part
of the calculations in the Hirn-Sanz model \cite{Hirn:2005nr} as a
check.

The organization of the paper is as follows. In the next section we
will briefly introduce the hard-wall AdS/QCD model, and review the
calculation of the $\gamma^*\gamma^*\pi^0$ form factor.
The extraction of $\gamma^*\rho^0\pi^0$ form factor and comparison
with other approaches will be presented in Sec.\,{I}{I}{I}. In
Sec.\,{I}{V} we give the $\gamma^*\rho^0\pi^0$ form factor in the
Hirn-Sanz model and compare it to that in the hard-wall model. The
last section is reserved for the summary.

\section{Extended AdS/QCD model with Chern-Simons
term}\label{sec:CS}
\subsection{hard-wall AdS/QCD model}
In the hard-wall model~\cite{Erlich:2005qh}, the background is given
by a slice of AdS space with the metric:
\begin{equation}
\mathd s^2=g_{MN}dx^Mdx^N = \frac{1}{z^2}\left(\eta_{\mu \nu}\mathd
x^{\mu}dx^{\nu} - \mathd z^2\right) ,
\end{equation}
where $ \eta_{\mu\nu} = {\rm Diag}\, (1,-1,-1,-1) $ and $ \mu, \nu =
(0,1,2,3) $, $ M, N = (0,1,2,3,z) $. The $\tmop{SU} \left( N_f
\right) \times \tmop{SU} \left( N_f \right)$ chiral symmetry is
realized through the gauge symmetry of two sets of gauge fields
$A_{(L)}$ and $A_{(R)}$. To breaking the chiral symmetry to the
vector part, an additional scalar field $X$ is introduced. The whole
action is then given by:
\begin{equation}
\label{AdS} S_{\rm AdS} = {\rm Tr}  \int \mathd^5x
 ~\left[\frac{1}{z^3}(D^{M}X)^{\dagger}(D_{M}X) +
\frac{3}{z^5} X^{\dagger}X -
\frac{1}{4g_5^2z}(F_{(L)}^{MN}F_{(L)MN}+F_{(R)}^{MN}F_{(R)
MN})\right] ,
\end{equation}
where $A=A^at^a, F_{MN}=\partial_M A_N-\partial_N A_M-i[A_M,A_N], D
X = \partial X - iA_{(L)}X + iX A_{(R)} $, and the generators are
normalized as $\mbox{Tr}~t^at^b=\delta^{ab}/2$. The vacuum solution
$\langle X(x,z)\rangle = v(z)/2= (m_q z + \sigma z^3)/2$ then breaks
the chiral symmetry to the vector part, and the phase of the
fluctuations of $X(x,z)$ gives the pion field: $X(x,z)= \langle
X\rangle\mbox{e}^{2i t^a \pi^a(x,z)}$. The vector combination
$V=(A_{(L)}+A_{(R)})/2$ corresponds to the vector mesons. Taking the
axial gauge $V_z=0$ and Fourier transforming to the 4D momentum
space, the transverse components $V_\mu^{\rm{T}}$ then satisfies the
following equation:
\begin{equation}
\partial_z\left(\frac{1}{z}\partial_zV^{\rm{T}}_\mu(q,z)\right)+\frac{q^2}{z}V^{\rm{T}}_\mu(q,z)=0.
\end{equation}
With Neumann boundary condition chosen at the cutoff, the normalized
solution is simply given by
\begin{equation}\label{holographicwf}
\psi_n^V ( z)   = \frac{\sqrt{2} }{z_0 J_1(\gamma_{0,n})}\, z
J_1(M_{n} z)
\end{equation}
with $ \gamma_{0,n}$ being the $n^{\rm th}$ zero of the Bessel
function $J_0(x)$ and $M_{n} = \gamma_{0,n}/z_0 $. Matching to the
experimental $\rho$ mass fixes $ z_0 = (323 \ {\rm MeV})^{-1} $. The
coupling constants, which are defined by $\langle
0|J^a_\mu|\rho_n^b\rangle=f_n\delta^{ab}\varepsilon_\mu$, can be
obtained by analyzing the two-point correlation function derived
from the action. The results are expressed through $\psi_n^V ( z)$
as
\begin{equation}
 f_n =\frac1{g_5} \,  \left  [ \frac1{z}\, \partial_z \psi^V_n (z) \right ]_{z=0} =
\frac{\sqrt{2}  M_n} {g_5z_0  J_1 (\gamma_{0,n})} \ .
\end{equation}
The non-normalized solution, or the bulk-to-boundary propagator, can
also be derived analytically:
\begin{equation}
\label{JQz} {\cal J} (Q,z) = {Qz}\left[K_1(Qz) + I_1(Qz)
\frac{K_0(Qz_0)}{I_0(Qz_0)} \right],
\end{equation}
where ${\cal J} (Q,z)$ is taken at a spacelike momentum $q$ with
$q^2 =-Q^2$ and satisfies the boundary condition ${\cal J}(Q,0)= 1$.
$I_n$ and $K_n$ are the order-$n$ modified Bessel functions of the
first and second kind, respectively. It can be shown that ${\cal J}
(Q,z)$ has the following decomposition
formula~\cite{Strassler2004,Radyushkin2007}:
\begin{equation}
\label{eq:Jdf}  {\cal J } (Q,z) = g_5\sum_{m = 1}^{\infty}\frac{
f_{m} \psi_m^V( z)}{ Q^2 + M^2_{m} }
\end{equation}
From ${\cal J } (Q,z)$ the vector current correlator can be derived,
whose asymptotic behavior determines the 5D coupling $g_5=2\pi$.

The axial combination $A=(A_{(L)}+A_{(R)})/2$ is a little
complicated because the longitudinal part will be entangled with the
chiral field. The equations are as follows:
\begin{equation}
  \partial_z\left(\frac1z \partial_z A^{\rm{T}}_\mu \right) + \frac{q^2}z
  A^{\rm{T}}_\mu
- \frac{g_5^2 v^2}{z^3} A^{\rm{T}}_\mu =0;\label{eq:AT}
\end{equation}
\vskip -1.5em
\begin{equation}
  \partial_z\left(\frac1z \partial_z \varphi \right)
+\frac{g_5^2 v^2}{z^3} (\pi-\varphi) = 0;\label{eq:varphi}
\end{equation}
\vskip -1.5em
\begin{equation}
  -q^2\partial_z\varphi+\frac{g_5^2 v^2}{z^2} \partial_z \pi =0.\label{eq:pi}
\end{equation}
where $\partial_\mu\varphi=A_\mu-A_{\mu}^{\rm{T}}$. The
normalization of $\varphi$ and $\pi$ is fixed by the pion kinetic
term:
\begin{equation}\label{eq:NC}
 \int_0^{z_0}\! \mathd z\, \left(\frac{\varphi'(z)^2}{g_5^2\, z} + \frac{v(z)^2
(\pi-\varphi)^2}{z^3} \right)=f_\pi^2.
\end{equation}
This normalization naturally leads to the charge conservation
constraint for the electromagnetic form factor of the pion, since
the pion form factor is given by~\cite{Kwee:2007dd,Grigoryan:2007wn}
\begin{equation}\label{eq:ffpi}
F_{\pi}(Q^2)=\frac{1}{f_\pi^2} \int_0^{z_0}\! \mathd z\, {\cal J }
(Q,z)\left(\frac{\varphi'(z)^2}{g_5^2\, z} + \frac{v(z)^2
(\pi-\varphi)^2}{z^3} \right).
\end{equation}
Notice that both the equations and the normalization condition is
invariant if the $\varphi$ and $\pi$ fields are shifted by a
constant simultaneously.

In the chiral limit $m_q=0$, the pion decay constant can be derived
from the residue of the axial current correlator at
$q^2=0$~\cite{Erlich:2005qh}:
\begin{equation}\label{eq:fpi-c}
f_\pi^2 = -\frac1{g_5^2}\left.\frac{\partial_z
A_{\rm{c}}(0,z)}{z}\right|_{z=\epsilon},
\end{equation}
where $A_{\rm{c}}(0,z)$ is the nonnormalizable solution to
Eq.~(\ref{eq:AT}) at $q^2=0$, satisfying $A_{\rm{c}}'(0,z_0)=0$ and
$A_{\rm{c}}(0,0)=1$. We use the subscript {``}c" to indicate that
the solution is obtained in the chiral limit. The explicit form of
$A_{\rm{c}}(0,z)$ is given by
\begin{equation}\label{eq:AT0}
A_{\rm{c}}(0,z) = {z\, \Gamma \left ({2}/{3} \right )
\left(\frac{\alpha}{2}\right)^{1/3}} \left[ I_{-1/3}\left(\alpha
z^3\right)  -  I_{1/3}\left(\alpha z^3\right)
\frac{I_{2/3}\left(\alpha z^3_0\right)} {I_{-2/3}\left(\alpha
z^3_0\right)}\right],
\end{equation}
where $\alpha\equiv g_5\sigma/3$. Matching to the experimental value
of $f_\pi$, one obtains $ \alpha = (424\,{\rm MeV})^3
$~\cite{Grigoryan:2008up}, or $ \sigma = (332\,{\rm MeV})^3 $. In
this case $\pi_{\rm{c}}(z)$ is just a constant and can be shifted to
zero. Then $\varphi_{\rm{c}}(z)$ satisfies the same equation as
$A_{\rm{c}}(0,z)$ with the same boundary condition at $z=z_0$, so we
have $\varphi_{\rm{c}}(z)=\varphi_{\rm{c}}(0) A_{\rm{c}}(0,z)$. The
normalization condition (\ref{eq:NC}) finally fixes
$\varphi_{\rm{c}}(0)$ to be one.

When $m_q\ne 0$, $A(q^2,z)$ will generally develop a $z\log z$ term,
unless $q^2=g_5^2 m_q^2$. This should not be identified with the
pion pole. Away from this point, $\partial_z A(q^2,z)/z$ is
divergent as $z\to 0$, which makes the generalization of
Eq.~(\ref{eq:fpi-c}) unfeasible. This can be overcome if we choose
$\varphi(z)$, rather than $A(0,z)$, to define the pion decay
constant. These two are identical in the chiral limit, but different
now. Thus we have
\begin{equation}\label{eq:fpi}
f_\pi^2 = -\frac1{g_5^2}\left.\frac{\partial_z
\varphi(z)}{z}\right|_{z=\epsilon}.
\end{equation}
One may worry that if this definition is consistent with the
normalization (\ref{eq:NC}). Notice that in this case
$(\varphi-\pi)$ is forced to vanish at the ultraviolet boundary.
Integration by parts and imposing the equation of motion with the
boundary condition, Eq.~(\ref{eq:NC}) becomes
\begin{equation}
\frac{m_\pi^2}{g_5^4}\int_0^{z_0}\frac{z}{v(z)^2}\varphi^{'}(z)^2\,\mathd
z=f_\pi^2,
\end{equation}
Using the same trick as in ref.~\cite{Erlich:2005qh} and
ref.~\cite{Abidin:2009aj}, one can show that the above condition
together with Eq.~(\ref{eq:fpi}) leads to the Gell-Mann-Oakes-Renner
relation.

\begin{figure}[htbp]
\centerline{$$\epsfxsize=0.9\textwidth\epsffile{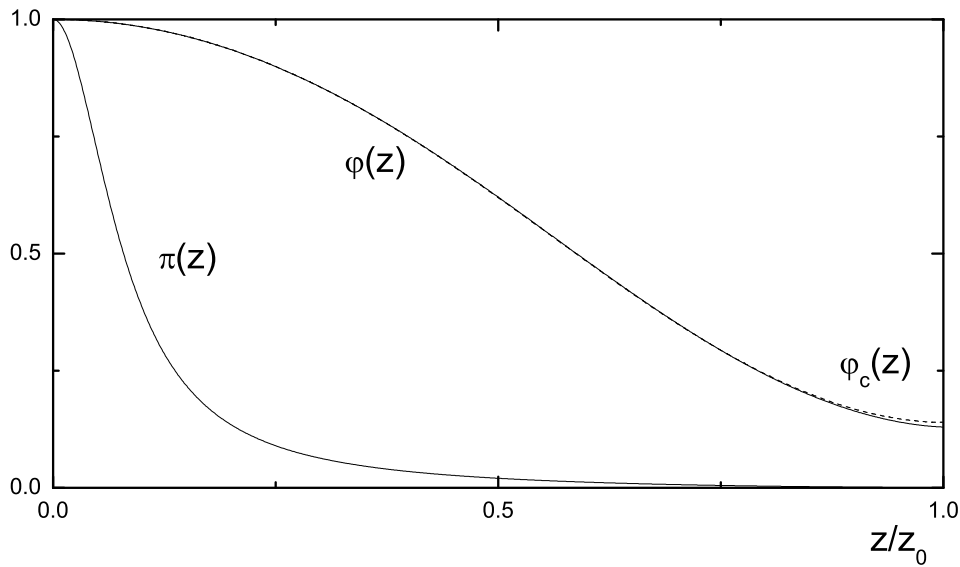}$$}
\caption{\it Explicit form of the solution $\varphi(z)$ and
$\pi(z)$~(solid curves), together with $\varphi_{\rm{c}}(z)$~(dashed
line), the solution in the chiral limit.}\label{fig:varphi-pi}
\end{figure}

\begin{figure}[htbp]
\centerline{$$\epsfxsize=0.9\textwidth\epsffile{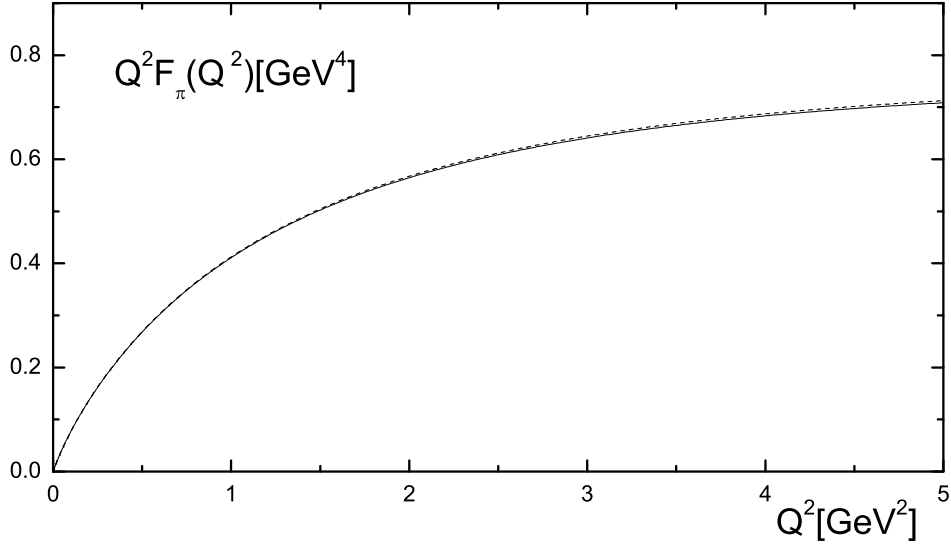}$$}
\caption{\it The pion electromagnetic form factor calculated in the
hard-wall model, the solid line denotes the result with finite quark
mass and the dashed one in the chiral limit.}\label{fig:FFpi}
\end{figure}

To solve the equations for $\varphi(z)$ and $\pi(z)$, one has to
employ numerical methods. We choose the boundary value
$\varphi(0)=\pi(0)=1$, for the convenience of comparing with the
chiral solution. Adjusting $m_\pi$ and $f_\pi$ to the experimental
value $m_{\pi_0}=135.0~\mbox{MeV}$ and $f_\pi=92.4~\mbox{MeV}$, one
obtains $m_q=2.14~\mbox{MeV}$ and $\sigma=(329~\mbox{MeV})^3$. The
explicit form of $\varphi(z)$ and $\pi(z)$ are plotted in
Fig.~(\ref{fig:varphi-pi}), together with the solution
$\varphi_{\rm{c}}(z)$. Surprisingly, the curves of $\varphi(z)$ and
$\varphi_{\rm{c}}(z)$ almost coincide, and one can hardly
distinguish them. $\pi(z)$ is also very close to
$\pi_{\rm{c}}(z)=0$, except for a peak near $z=0$. Thus one may
expect that the quark mass correction to any physical observable
would be small. To check this, we recalculate the pion form factor
according to Eq.~(\ref{eq:ffpi}), which has already been done in
ref.~\cite{Kwee:2007dd} for finite quark mass, and in
ref.~\cite{Grigoryan:2007wn} in the chiral limit. The result
confirmed our expectation, see Fig.~(\ref{fig:FFpi}). We also
calculated some other observable and obtained similar results in
both cases.

\subsection{Extended hard-wall model with Chern-Simons term}
Now let us turn to the derivation for the $\gamma^*\gamma^*\pi^0$
form factor by adding a CS term to the hard-wall model, pioneered by
Grigoryan and Radyushkin~\cite{Grigoryan:2008up}. First we should
enlarge the previous considered $SU(2)_L  \otimes SU (2)_R$ gauge
group into $U(2)_L \otimes U(2)_R $. To do this, we replace the
gauge fields $t^a A^a_\mu$ in the action by ${\cal A}_{\mu} = t^a
A^a_\mu + \mbox{I} \, \frac{{\hat A}_\mu }{2}$. The gauge field
$\hat A_\mu$ will couple to the isosinglet current in the boundary
theory. The cubic part of the CS term can be expressed in the axial
gauge as
\begin{equation}
\label{cs} S^{(3)}_{\rm CS}[{\cal A}] = k\,
\frac{N_c}{48\pi^2}\epsilon^{\mu\nu\rho\sigma} {\rm Tr} \int
\mathd^4 x\, \mathd z  \left(\partial_z {\cal A}_{\mu}\right)
\biggl[{\cal F}_{\nu \rho}{\cal A}_{\sigma} + {\cal A}_{\nu}{\cal
F}_{\rho \sigma} \biggr],
\end{equation}
with $k$ an integer. For the $U(2)_L \otimes U(2)_R $ gauge group,
the corresponding cubic action reads:
\begin{equation}
S^{\rm AdS}_{\rm CS}[{\cal A}_{(L)}, {\cal A}_{(R)}] = S^{(3)}_{\rm
CS}[{\cal A}_{(L)}]  - S^{(3)}_{\rm CS}[{\cal A}_{(R)}].
\end{equation}
The relevant term for the anomalous $\pi^0 \gamma^* \gamma^*$ form
factor can be found to be
\begin{equation}\label{actionPhi3}
S^{\rm  anom}  =k\, \frac{N_c}{8\pi^2}\epsilon^{\mu\nu\rho\sigma}
\int \mathd^4 x
  \int_0^{z_0}   dz \,  (\partial_z  \varphi^a) \left( \partial_\rho  V^a_{\mu} \right)
  \left(\partial_{\sigma}\hat{V}_{\nu}\right).
\end{equation}
Based on the holographic dictionary, one then obtains the bare form
factor as
\begin{equation}
F_{\gamma^* \gamma^* \pi^0}(Q_1^2,Q_2^2) = - \frac{N_c}{12 \pi^2
f_\pi} \cdot \frac{k}{2}   \int_0^{z_0}  {\cal J}(Q_1,z) {\cal
J}(Q_2,z)
\partial_z \varphi (z) \, \mathd z.
\end{equation}
In QCD, the axial anomaly determines the value of the form factor
with real photons to be $F_{\gamma^* \gamma^* \pi^0}(0,0)
=\frac{N_c}{12 \pi^2 f_\pi}$. To reproduce this result, a surface
term must be added, and the integer $k$  must be taken to be $2$.
The final result for the normalized function $K (Q_1^2,Q_2^2)$ is
then
\begin{eqnarray}
\label{kqqmod} K (Q_1^2,Q_2^2)  &=& F_{\gamma^* \gamma^*
\pi^0}(Q_1^2,Q_2^2)/F_{\gamma^* \gamma^* \pi^0}(0,0) \nonumber\\
&=&\varphi (z_0) {\cal J}(Q_1,z_0) {\cal J}(Q_2,z_0) -  \int_0^{z_0}
{\cal J}(Q_1,z) {\cal J}(Q_2,z)\,   \partial_z \varphi (z) \, \mathd
z.
\end{eqnarray}
This result has very interesting properties~\cite{Grigoryan:2008up}.
When one photon is real, the form factor has the following expansion
at low momentum :
\begin{equation}
K(0,Q^2) = 1- a_\pi \, \frac{Q^2}{m_{\pi}^2},
\end{equation}
with $a_\pi \approx 0.031$ in perfect agreement with the
experimental value: $a|_{\rm exp}\simeq 0.032\pm 0.004$. This
indicates a strong Vector Meson Dominance~(VMD) in this channel,
which will lead to the result $a=\frac{m^2_\pi}{m^2_{\omega}}\simeq
0.03$. At large virtuality for one or both photons, the asymptotic
behavior of the form factor can be found to be:
\begin{eqnarray}
K(0,Q^2)         &\to&  \bar s/Q^2,\nonumber\\
K(Q_1^2, Q_2^2)  &\to&  \frac{\bar s}{3Q^2} \int_0^1 \frac{6 \, x
(1-x) \, dx}{1+\omega (2x-1)},
\end{eqnarray}
where $\bar s=8\pi^2f_\pi^2$. Both coincide with the leading-order
pQCD results calculated for the asymptotic form of the pion
distribution amplitude. However, the origins of the power behavior
are quite different. The power behavior appears only after we have
integrated out the meson wave function in the holographic direction.
This is very similar to the {``}soft" contributions described in the
LCSR approach, which will be discussed in the following section.

\section{Form factor of $\gamma^{*}\rho^0\to\pi^0$ transition}
It will be illuminating to further study the $\rho^0\to\pi^0$
transition form factor based on the previous result. One starts with
the dispersion relation for the amplitude $F_{\gamma^* \gamma^*
\pi^0}(Q_1^2,Q_2^2)$ in the variable $Q_2^2$ and at fixed $Q_1^2$.
In the standard QCD sum rule approach, one assumes that the spectral
density in the dispersion relation can be approximated by the ground
states $\rho^0$, $\omega$ and the the higher states with an
effective threshold $s_0$:
\begin{equation}
 F^{\gamma^{*}\gamma^{*}\pi^0}(Q_1^2,Q_2^2) =\frac{\sqrt{2}
f_\rho F^{\rho^0 \pi^0}(Q_1^2)}{m_{\rho^0}^2+Q_2^2} +
\int\limits_{s_0}^\infty \mathd s~
\frac{\rho^h(Q_1^2,s)}{s+Q_2^2}.\label{disp}
\end{equation}
Here, the $\gamma^{*}\rho^0(\omega)\to\pi^0$ form factor is defined
as
\begin{equation}
 \frac{1}{3}\,\langle \pi^0\mid j_\mu(q_1)\mid \omega(q_2)\rangle
= \langle \pi^0\mid j_\mu(q_1)\mid \rho^0(q_2)\rangle = F^{\rho^0
\pi^0}(Q_1^2)
m_\rho^{-1}\epsilon_{\mu\nu\alpha\beta}e^{\nu}q_1^\alpha q_2^\beta,
\label{formf}
\end{equation}
and the decay constants have the relation:
\begin{equation}
3\,\langle \omega \mid j_\mu \mid 0 \rangle = \langle \rho^0 \mid
j_\mu \mid 0 \rangle = \frac{f_\rho}{\sqrt{2}} m_\rho e^{
*}_\nu, \label{formrho}
\end{equation}
$e_\mu$ being the polarization vector of the $\rho(\omega)$ meson.
Since we are working in the $U(2)_{\rm{V}}$ symmetric limit, the
above relations are exact. On the other hand, the dispersion
relation can be carried out explicitly using the decomposition
formula~(\ref{eq:Jdf}). Extracting the lowest $\rho^0$ and $\omega$
pole contributions, one immediately obtains the expression for the
$\gamma^*\rho_0\pi_0$ form factor:
\begin{equation}
F^{\rho^0\pi^0}(Q^2)=\frac{N_c}{12\pi^2 f_\pi}
\frac{g_5m_\rho}{2}\left[ {\cal J}(Q,z_0) \,\psi_1^V (
z_0)\,\varphi(z_0)-\int_0^{z_0}{\cal J}(Q,z) \,\psi_1^V (
z)\,\partial_z\varphi(z)\mathd z\right] \label{eq:ff}
\end{equation}
As discussed before, the quark mass correction is very small, so we
mainly work in the chiral limit. The results for finite quark mass
are listed only in Table~\ref{tab:dw} for comparison.

\subsection{Large $Q^2$ region}
First let us focus on the large-$Q^2$ asymptotic behavior of the
form factor. The large-$Q^2$ behavior of ${\cal J} (Q,z)$ is
dominated by the term $ zQ K_1 (zQ)$ in Eq.~(\ref{JQz}), which
behaves like $e^{-Qz}$. Thus the first term in Eq.~(\ref{eq:ff})
will vanish exponentially $\sim e^{-Qz_0}$ in the asymptotic region,
hence can be neglected.  Due to the exponential factor of ${\cal J}
(Q,z)$, only small values of $z$ are important in the remaining
integral, and the outcome is determined by the small-$z$ behavior of
the wave function $\partial_z\varphi(z)$ and $\psi_1^V ( z)$. From
the previous discussion, we know that when $z\to 0$,
\begin{equation}
\partial_z\varphi(z)\sim -f_\pi^2g_5^2z,
\end{equation}
and
\begin{equation}
\psi_1^V ( z)\sim \frac{m_\rho z^2}{\sqrt{2}z_0J_1(\gamma_{0,n})}.
\end{equation}
Utilizing all these facts one finds:
\begin{eqnarray}
F^{\rho^0\pi^0}(Q^2)&\to& \frac{N_cm_\rho}{12\pi f_\pi}
\frac{m_{\rho}}{\sqrt{2}z_0J_1(\gamma_{0,1})}(-f_\pi^2g_5^2)
\int_0^{z_0} z^3* zQK_1(zQ)\mathd z \, \nonumber\\
&=&\frac{\pi f_\pi m_\rho^3}{\sqrt{2}\gamma_{0,1}J_1(\gamma_{0,1})}\frac{\int_0^\infty \chi^4K_1(\chi) d\chi}{Q^4}\nonumber\\
&=&\frac{8\sqrt{2}\pi f_\pi
m_\rho^3}{\gamma_{0,1}J_1(\gamma_{0,1})}\frac{1}{Q^4}=\frac{1.23\,\mbox{GeV}^4}{Q^4}
\end{eqnarray}
Using the holographic expression for $m_\rho$ and $f_\rho$, one may
further express the result as
\begin{equation}
F^{\rho^0\pi^0}(Q^2) \to \frac{8\sqrt{2}\pi^2f_\pi f_\rho
m_\rho^2}{Q^4}. \label{eq:asff1}
\end{equation}
Although this power behavior is the same as the pQCD prediction
\cite{Chernyak:1983ej}, the underlying mechanism is rather
different. The appearance of the power behavior in this way is
similar to the LCSR
analysis~\cite{Braun:1994ij,Khodjamirian:1997tk}, where the form
factor is given by the following expression:
\begin{eqnarray}
F_{\rm{LC}}^{\rho\pi}(Q^2) &=&
\frac{f_\pi}{3f_\rho}\int^1_{\frac{Q^2}{s_0+Q^2}}\!\frac{\mathd u}u
\Bigg(\varphi_\pi
(u)+\frac{u}{Q^2}\frac{\mathd\varphi^{(4)}(u)}{\mathd u}\Bigg)
\exp\left(-\frac{Q^2(1-u)}{uM^2}
+\frac{m_\rho^2}{M^2}\right)\label{eq:LCff}\\
&=&\frac{f_\pi}{3f_\rho}\frac{\varphi_\pi
'(1)}{Q^4}\exp\left(\frac{m_\rho^2}{M^2}\right)\int_0^{s_0}\!s\mbox{e}^{-s/M^2}\mathd
s+\mbox{O}(1/Q^6).
\end{eqnarray}
Here $\varphi_\pi (u)$ and $ \varphi^{(4)}(u)$ are the leading twist
and the twist-4 distribution amplitudes, and $M^2$ the Borel
parameter. In deriving the asymptotic behavior, we have assumed that
$\phi_\pi(u) \stackrel{u\rightarrow 1}{\sim} \varphi_\pi '(1)(1-u)$.
That is to say, in order to obtain the same power behavior as in the
holographic model, $\phi_\pi(u)$ must has the same end-point
behavior as the asymptotic one, namely
$\phi_\pi^{\rm{as}}(u)=6u(1-u)$. This is also in accordance with the
general analysis for the end-point behavior of the pion wave
function~\cite{Chernyak:1983ej}.

Moreover, in the holographic approach ${\cal J} (Q,z)\to e^{-Qz}$
tells us in the large $Q^2$ limit we are actually probing the
$0<z<1/Q$ interval of the AdS slice, while in the LCSR
(\ref{eq:LCff}) the endpoint region, $0<1-u<1/Q^2$, or
$0<\sqrt{1-u}<1/Q$ dominates. Taking into account the symmetry
$u\leftrightarrow 1-u$ of the light quark system, one may expect
that $ z $ should be dual to $\sqrt{u(1-u)}b$ with $b$ a light-cone
distance parameter, at least in the high energy region. This is just
one of the key relations of the Light-Front
holography~\cite{Brodsky:2006uqa,Brodsky:2007hb}.

\begin{figure}[htbp]
\centerline{$$\epsfxsize=0.9\textwidth\epsffile{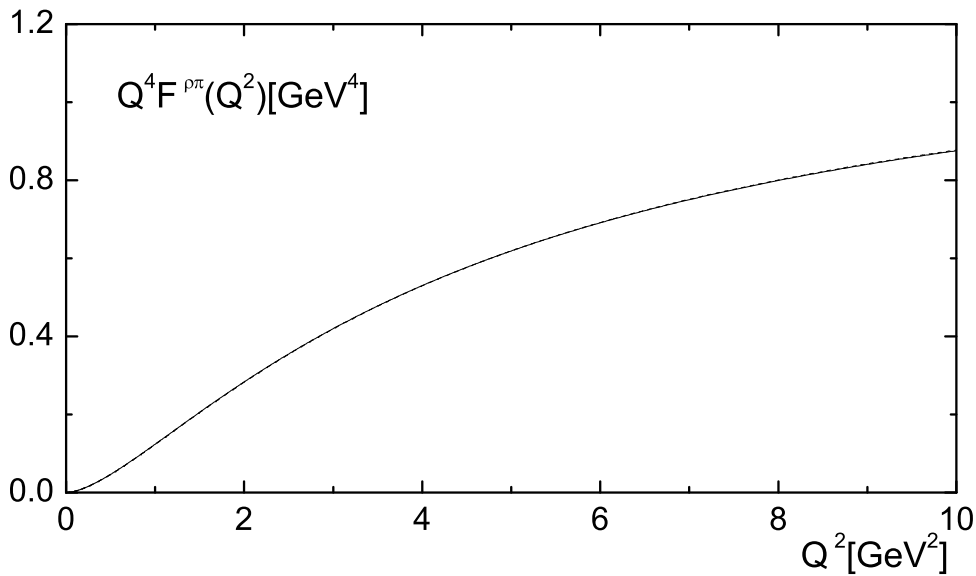}$$}
\caption{\it $\gamma^*\rho^0 \to \pi^0$ form factor calculated in
the extended hard-wall AdS/QCD model, the result for finite quark
mass~(in solid curve) and that in the chiral limit (dashed line)
almost coincide.}\label{fig:FFrhopi}
\end{figure}

\begin{figure}[htbp]
\centerline{
\epsfig{bbllx=79pt,bblly=256pt,bburx=575pt,%
bbury=550pt,file=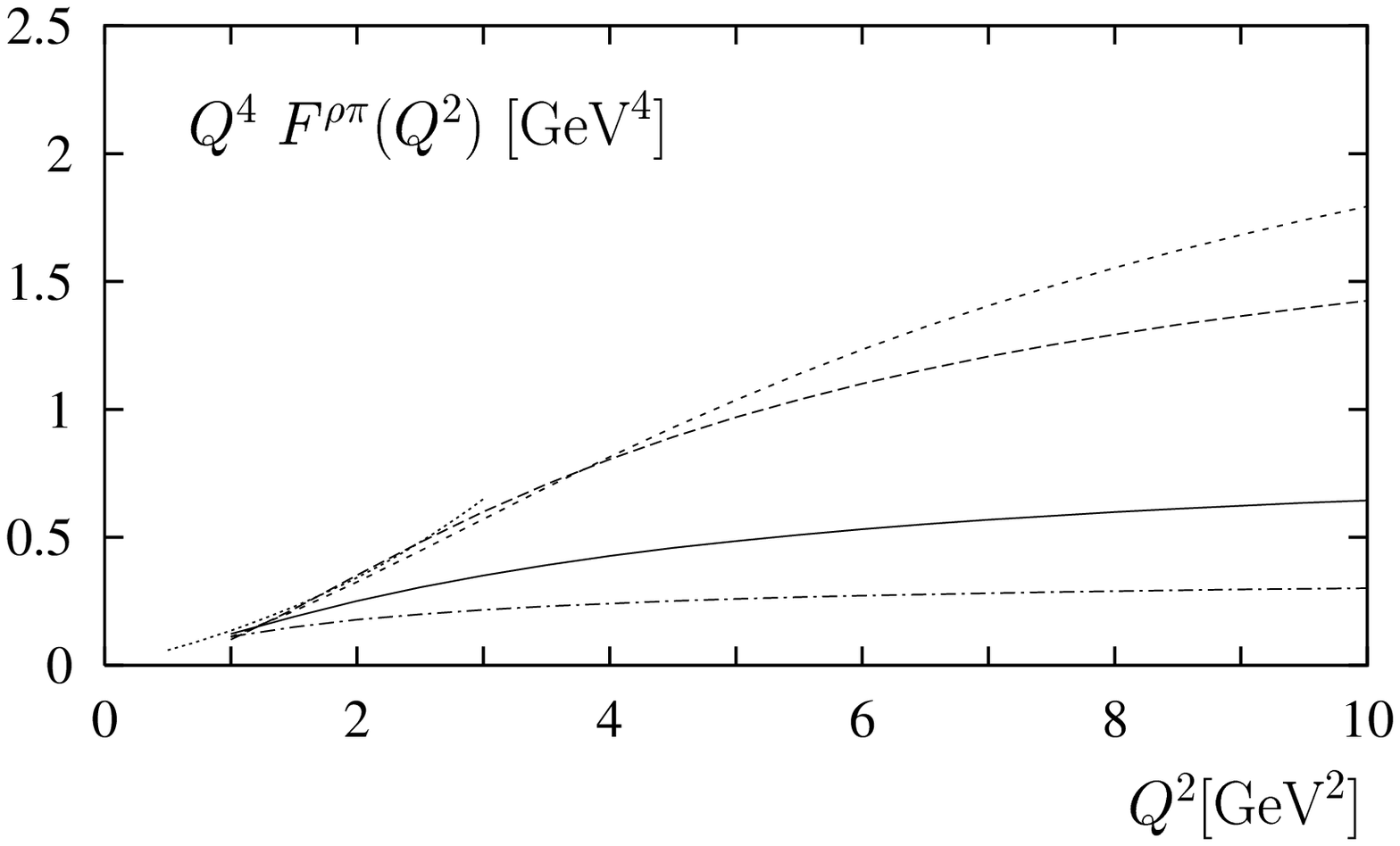,scale=0.8,%
,clip=} } \caption{\it LCSR results for $\gamma^*\rho^0 \to \pi^0$
form factor excerpted from \cite{Khodjamirian:1997tk}. The solid
line corresponds to the result calculated with the asymptotic pion
wave function, while the long-dashed and the short-dashed lines with
the CZ and BF wave functions respectively. In comparison, the
predictions of the three-point QCD sum rule (dotted)
\cite{Eletsky:1982im} and an alternative light-cone sum rule for the
$\gamma^*\rho^0\to \pi^0$ form factor \cite{Braun:1994ij}
(dash-dotted) were also plotted.}\label{fig:FFlcsr}
\end{figure}

Since the absolute normalization of the asymptotic behavior is not
known in both the pQCD and LCSR approaches, one can only compare
their predictions to the form factor at moderate $Q^2$ with ours. At
$Q^2\simeq 10\,\mbox{GeV}^2$, direct calculation from
Eq.~(\ref{eq:ff}) gives $F^{\rho^0\pi^0}(Q^2)=8.8\times10^{-3}$, as
shown in Fig.~\ref{fig:FFrhopi}. This is much larger than the pQCD
result
$F^{\rho\pi}_{\rm{pQCD}}(Q^2)\simeq3\times10^{-3}$~\cite{Chernyak:1983ej}.
In the LCSR approach, the result strongly depends on the shape of
the leading twist distribution amplitude of the pion meson, which
can be seen from Fig.~\ref{fig:FFlcsr}. For the asymptotic
distribution amplitude, one obtains
$F^{\rho\pi}_{\rm{as}}(Q^2)\simeq 6.6\times10^{-3}$~\footnote{In
Ref.~\cite{Braun:1994ij} an alternative light-cone sum rule was
derived for this form factor, from which a much smaller value
$F^{\rho\pi}_{\rm{as}}(Q^2\simeq~10\mbox{GeV}^2)\simeq
3\times10^{-3}$ (see Fig.~\ref{fig:FFlcsr}) was obtained. However,
with the aid of the technique in Ref.~\cite{Ball:1997rj} one can
show that a boundary term was missing in their calculations. After
including this term, a similar result
$F^{\rho\pi}_{\rm{as}}(Q^2)\simeq 7.0\times10^{-3}$ will be
obtained. }. For some non-asymptotic distribution amplitudes, the
results are much larger, e.g., input of the
Chernyak-Zhitnitsky~(CZ)~\cite{Chernyak:1983ej} and
Braun-Filyanov~(BF)~\cite{Braun:1988qv} distribution amplitudes
give: $F^{\rho\pi}_{\rm{CZ}}(Q^2)\simeq 0.014$,
$F^{\rho\pi}_{\rm{BF}}(Q^2)\simeq 0.017$, respectively. Thus our
result indicates that the true pion distribution amplitude should be
asymptotic-like, in accordance with the conclusion made from the
studies of the $\gamma^*\gamma^*\pi^0$ form
factor~\cite{Grigoryan:2008up}.


\subsection{Low $Q^2$ region}
The $\gamma\rho^0\pi^0$ form factor at zero momentum transfer
defines the coupling constant g$_{\rho^0\pi^0\gamma}$ through the
effective Lagrangian~\cite{Titov:1999eu}
\begin{equation}
{\cal
L}^{eff}_{\rho\pi\gamma}=g_{\rho\pi\gamma}m_\rho^{-1}\varepsilon_{\mu\nu\alpha\beta}
\partial^\mu \rho^{0\nu} \partial^\alpha A^\beta\pi^0 .\label{eq:eL}
\end{equation}
Since ${\cal J}(Q,0) = {\cal J}(0,z) = 1$, one immediately obtains
the coupling constant:
\begin{eqnarray}
g_{\rho^0\pi^0\gamma}&=&\frac{N_cm_\rho}{12\pi f_\pi}\left[\psi_1^V
( z_0)\,\varphi(z_0)-\int_0^{z_0}{\psi_1^V (
z)\,\partial_z\varphi(z)\mathd z}\right]\nonumber\\
&=& 0.56.
\end{eqnarray}
This result is very close to the value extracted from the analysis
of $\rho^0$ and $\omega$ photoproduction reactions through
pseudoscalar exchange, which gives rise to
$g_{\rho\pi\gamma}=0.54$~\cite{Oh:2000pr} . Also it is consistent
with the QCD sum rule prediction $g_{\rho\pi\gamma}=0.63\pm
0.07$~\cite{Gokalp:2001sr}. Based on the effective Lagrangian given
in Eq.~(\ref{eq:eL}), the decay width for $V\rightarrow\pi^0\gamma$
can be readily deduced:
\begin{equation}\label{e6}
  \Gamma(V\rightarrow\pi^0\gamma)=\frac{\alpha}{24}
  \frac{(m_V^2-m_{\pi^0}^2)^3}{m_V^5}g_{V\pi\gamma}^2.
\end{equation}
Substituting the physical masses of the mesons and taking
$\alpha=1/137$, the partial widths
$\Gamma(\rho^0\rightarrow\pi^0\gamma)$ and
$\Gamma(\omega\rightarrow\pi^0\gamma)$ can be obtained.
We can further extrapolate the form factor to the time-like region
by analytically continuing ${\cal J}(Q,z)$ to the region
$q^2=-Q^2>0$,
\begin{equation}
{\cal
J}(q,z)=-\frac{\pi}{2}qz\left[Y_1(qz)-J_1(qz)\frac{Y_0(qz)}{J_0(qz)}\right]
\end{equation}
with $Y_n$ the second kind Bessel function.
From the resulting $\gamma^*\rho^0(\omega)\pi^0$ form factor in the
time-like region, one obtains the decay widths for the
$\rho^0(\omega)\to \pi^0 e^+e^-$ and $\rho^0(\omega)\to \pi^0
\mu^+\mu^-$ decays. In Table~\ref{tab:dw} we list these results
together with those for the radiative decays. The only reason we
keep four digits for our predictions is to show the corrections due
to finite quark mass.
\begin{table}[tp]
\small
\begin{center}
\begin{tabular}{llllll}
\hline
   & {\rm Experiment~\cite{Amsler:2008zzb}}
     &{$m_q=0$}  &{$m_q\ne 0$}
\\ \hline
  $\Gamma(\rho^0\rightarrow \pi^0\gamma)$ & $0.090\pm0.013$   &
  $0.06735$  &$0.06740$
  \\
  $\Gamma(\omega\rightarrow \pi^0\gamma)$ & $0.76\pm0.03$   &
  $0.6125$  & $0.6129$
             \\
  $\Gamma(\rho^0\rightarrow \pi^0 e^+ e^- )$ & ------   &
$6.167\times 10^{-4}$ & $6.172\times 10^{-4}$
            \\
  $\Gamma(\rho^0\rightarrow \pi^0\mu^+\mu^-)$ & ------   &
$6.422\times 10^{-5}$  &$6.427\times 10^{-5}$
  \\
  $\Gamma(\omega\rightarrow \pi^0  e^+ e^-  )$ & $(6.5\pm0.8)\times10^{-3}$   &
$5.629\times 10^{-3}$ & $5.634\times 10^{-3}$
            \\
  $\Gamma(\omega\rightarrow \pi^0\mu^+\mu^-)$ & $(8.2\pm2.0)\times10^{-4}$   &
$6.015\times 10^{-4}$  & $6.019\times 10^{-4}$
  \\
\hline
\end{tabular}
\caption{\it Predictions of the partial decay widths (in MeV) of
$\rho^0$ and $\omega$ in the present approach, both in the chiral
limit and with finite quark mass.} \label{tab:dw}
\end{center}
\end{table}

\section{$\gamma^*\rho^0\pi^0$ Form factor in the extended Hirn-Sanz Model}
Spontaneous chiral symmetry breaking can also be implemented through
the boundary conditions at the IR cutoff, without employing the
scalar field, as proposed by Hirn and Sanz~\cite{Hirn:2005nr}.
Specifically, the axial combination of the left-handed and
right-handed vector fields was chosen to satisfy Dirichlet boundary
condition, rather than the Neumann one for the vector part. That is
to say, we require:
\begin{eqnarray}
  F_{(R)z \mu} \left( x, z = z_0 \right) + F_{(L)z \mu} \left( x, z = z_0 \right) &
  = & 0 .  \label{R5+L5IR}\\
  A_{(R)\mu} \left( x, z = z_0 \right) - A_{(L)\mu} \left( x, z = z_0 \right) & = &
  0 .  \label{R-LIR}
\end{eqnarray}
Then the 5D gauge transformations for $R_{\mu}$ and $L_{\mu}$ at the
point $z = z_0$ must be equal. The chiral field can be defined as
\begin{eqnarray}
  U ( x ) & \equiv & \xi_R \left( x, z_0 \right) \xi_L^{\dag} \left( x, z_0
  \right),
\end{eqnarray}
where the Wilson line is defined as
\begin{eqnarray}
  \xi_{R(L)} \left( x, z \right) & \equiv & \text{P} \left\{ \mathe^{\mathi
  \int_{z_0}^z \mathd z A_{(R)z}(A_{(L)z}) \left( x, z \right) \bignone} \right\} ,
\end{eqnarray}
with $\text{P}$ denoting path-ordered integral. The equality of the
5D gauge symmetry at $z = z_0$ enforces the following transformation
law for the chiral field
\begin{eqnarray}
  U \left( x \right) & \longmapsto & g_R \left( x \right) U \left( x \right)
  g_L^{\dag} \left( x \right) .  \label{U-trsf}
\end{eqnarray}
where $\left( g_R, g_L \right)$ represent the 5D gauge symmetries
located on the UV brane, which are then interpreted as the 4D
$\tmop{SU} \left( N_f \right) \times \tmop{SU} \left( N_f \right)$
chiral symmetry. A vacuum state with $U=1$ naturally leads to the
spontaneous breaking of the symmetry group to the vector part.

To separate the dynamical fields and external sources from $A_{(L)}$
and $A_{(R)}$, one should first make a gauge transformation using
the above Wilson lines:
\begin{eqnarray}
  \hat{V}_M, \hat{A}_M & \equiv & \mathi \left\{ \xi^{\dag}_L  \left(
  \partial_M - \mathi A_{(L)M} \right) \xi_L \pm \left(L\to R \right) \right\} .  \label{hatted}
\end{eqnarray}
After making this transformation we are then working in the axial
gauge, $\hat{V}_z = \hat{A}_z = 0$. For the vector part, one can
simply make the following substraction
\begin{eqnarray}
  V_{\mu} \left( x, z \right) & \equiv & \hat{V}_{\mu} \left( x, z \right) -
  \hat{V}_{\mu} \left( x, z_0 \right),  \label{Vmu}
\end{eqnarray}
and the dynamics of $V_{\mu} \left( x, z \right)$ is completely the
same as in the original hard-wall model. However, to remove the
effect of the UV source of the axial field on the IR, a function
$\alpha(z)$ has to be introduced with the boundary values
\begin{equation}
\alpha(0)=1,~\alpha(z_0)=0.
\end{equation}
Subsequently, the axial field can be decomposed as
\begin{eqnarray}
  A_{\mu} \left( x, z \right) & \equiv & \hat{A}_{\mu} \left( x, z \right) -
  \alpha \left( z \right)  \hat{A}_{\mu} \left( x, z_0 \right) .  \label{Amu}
\end{eqnarray}
Since $\hat{A}_{\mu} \left( x, z_0 \right)$ contains the derivative
of the chiral field $U$, $\alpha \left( z \right) $ will play the
role of the 5D wave function of the pion. Moreover, to eliminate the
mixing of the dynamical axial field and the pion, $\alpha \left( z
\right) $ must satisfy
\begin{align}\label{seom}
\partial_z \left ( \frac{1}{z}\,  \partial_z\alpha  \right )  = 0  \ .
\end{align}
Together with the aforementioned boundary conditions, this fixes
$\alpha \left( z \right) $ to be of the form
\begin{eqnarray}
  \alpha \left(z\right) & = & 1 - {z^2}/{z_0^2}.
\end{eqnarray}
Substituting these decompositions into the original action, one can
naturally deduce the chiral lagrangian, with all the low energy
constants given by simple integrals of $\alpha \left( z \right) $.
Most importantly, one has
\begin{align}\label{fpi2}
f^2_{\pi} &= \frac{1}{g^2_5}\int^{z_0}_{0}\frac{\mathd z}{z}
\left(\partial_z \alpha\right)^2 = \frac{2}{g^2_5z^2_0}.
\end{align}
If $g_5$ and $z_0$ were fixed as before, we would have $ f_{\pi}
\simeq 72.7 \ \MeV $, which is somewhat smaller than the
experimental value. Due to this drawback, we will only repeat some
of the previous calculations in this model.

Again we start from the $\gamma^*\gamma^*\pi^0$ form factor in this
model, which has already been derived in
Ref.~\cite{Grigoryan:2008cc} along the same line as in the hard-wall
model:
\begin{eqnarray} \label{K12}
F_{\gamma^*\gamma^*\pi^0} \left(Q_1^2,Q_2^2 \right ) = -
\frac{N_c}{12 \pi^2 f_\pi} \int_0^{z_0} {\cal J}(Q_1,z) {\cal
J}(Q_2,z)\,
\partial_z \alpha(z) \, \mathd z,
\end{eqnarray}
where the normalization constant $k$ of the CS term has also been
chosen to be $2$. This is enough to ensure the anomaly relation
since $\alpha(z_0)=0$. No surface term at the IR boundary needs to
be introduced. From the above expression we see that $\alpha(z)$
indeed plays the role of pion wave function, as $\Psi(z)$ does in
the original hard-wall model. Moreover, the behavior of these two
functions near the UV boundary are also the same, since
\begin{equation}
\partial_z \alpha(z)=-2z/z_0^2=-f_{\pi}^2g_5^2z.
\end{equation}
From this one can conclude that the asymptotic behavior of the
$\gamma^*\gamma^*\pi^0$ form factor must be the same as in the
hard-wall model, which was found in Ref.~\cite{Grigoryan:2008cc}.

The $\gamma^*\rho^0\pi^0$ form factor can derived as in previous
sections, which is given by
\begin{equation}
F^{\rho^0\pi^0}(Q^2)=-\frac{N_c}{12\pi^2 f_\pi}
\frac{g_5m_\rho}{2}\left[ \int_0^{z_0}{\cal J}(Q,z) \,\psi_1^V (
z)\,\partial_z \alpha(z)\mathd z\right]. \label{eq:ff2}
\end{equation}
For the same reason as preceding discussion, its asymptotic behavior
is the same as Eq.~(\ref{eq:asff1}). The $\rho^0\pi^0\gamma$
coupling can also be obtained
\begin{equation}
g_{\rho^0\pi^0\gamma}=-\frac{N_c}{12\pi^2 f_\pi}
\frac{g_5m_\rho}{2}\int_0^{z_0}{\psi_1^V (
z)\,\partial_z\alpha(z)\mathd z}.
\end{equation}
Substituting the experimental value of $f_\pi$ in the normalization
factor, we get $g_{\rho^0\pi^0\gamma}=0.65$, in reasonable agreement
with the hard-wall result and those derived from other approaches.
In ref.~\cite{Domokos:2009cq} an exhaustive list of the three-point
and four-point couplings was given for the Hirn-Sanz model. The
corresponding value for the $\rho\pi\gamma$ vertex is
$g_{\rho\pi\gamma}=0.06~f_\pi^{-1}~\mbox{GeV}\simeq~0.649$,
confirming our result. Similar result was also obtained in
refs.~\cite{Pomarol:2008aa,Becciolini:2009fu}.

\section{Summary}
In this work, the $\gamma^*\rho^0 \to \pi^0$ transition form factor
has been extracted from the $\gamma^*\gamma^*\pi^0$ form factor,
which has been obtained in the extended hard-wall AdS/QCD model
including a Chern-Simons term. As expected from pQCD, the form
factor exhibits the $1/Q^4$ asymptotic behavior, but with a rather
different mechanism. It comes out only after we integrate the meson
solution with the bulk-to-boundary propagator along the holographic
direction. The power is then determined by the $z\to 0$ behavior of
the meson solution. The appearance of this power behavior is very
similar to that in the LCSR approach, where the power is dictated by
the end-point behavior of the Light-Cone distribution amplitude.
Comparing the corresponding expressions, one can deduce the dual
relation $z= \sqrt{u(1-u)}b$ with $b$ a light-cone distance
parameter, which is just one of the important relations in the
Light-Front holography. Since the numerical results of the form
factor in the LCSR approach strongly depend on the profile of the
pion distribution amplitude $\phi_{\pi}(u)$, the present analysis
can help to discriminate between various models for $\phi_{\pi}(u)$.
As in the discussion for the $\gamma^*\gamma^*\pi^0$ form factor,
our result favors an asymptotic-like pion distribution amplitude.
From the form factor at $Q^2=0$ we obtains the partial width of the
radiative decays $\rho^0(\omega)\to \pi^0\gamma$. We also extend our
analysis by analytically continuing the bulk-to-boundary propagator
to the time-like region. The Dalitz decays $\rho^0(\omega)\to \pi^0
e^+e^-, \pi^0\mu^+\mu^-$ are then studied. All these decay rates are
roughly consistent with the available measured values. The quark
mass corrections are found to be very small, as expected.

Some of the calculations have been performed in the Hirn-Sanz model,
which successfully describes the spontaneous chiral symmetry
breaking in a simple way. Just as in the case of the
$\gamma^*\gamma^*\pi^0$ form factor, the asymptotic behavior of the
$\gamma^*\rho^0\pi^0$ form factor in this model is exactly the same
as in the standard hard-wall model. The $\gamma\rho^0\pi^0$ coupling
is also in reasonable agreement with the hard-wall result.

\hspace{1cm}

{\bf Acknowledgments}: This work was supported in part by Natural
Science Foundation of China under Grant No.~10875130, No.~10935012,
No.~10805082, and No.~10675132. We would like to thank Andrea Wulzer
for good comments.


\begin{thebibliography}{10}

\bibitem{Maldacena1998}
Juan~Martin Maldacena.
\newblock Adv. Theor. Math. Phys. \textbf{2} (1998): 231-252
  [\href{http://arxiv.org/abs/hep-th/9711200}{arXiv: hep-th/9711200}].

\bibitem{Witten1998a}
Edward Witten.
\newblock Adv. Theor. Math. Phys. \textbf{2} (1998): 253-291
  [\href{http://arxiv.org/abs/hep-th/9802150}{arXiv: hep-th/9802150}].

\bibitem{Polyakov1998}
S.~S. Gubser, Igor~R. Klebanov, and Alexander~M. Polyakov.
\newblock Phys. Lett. \textbf{B428} (1998): 105-114
  [\href{http://arxiv.org/abs/hep-th/9802109}{arXiv: hep-th/9802109}].

\bibitem{Polchinski:2001tt}
Joseph Polchinski and Matthew~J. Strassler.
\newblock Phys. Rev. Lett. \textbf{88} (2002): 031601
  [\href{http://arxiv.org/abs/hep-th/0109174}{arXiv: hep-th/0109174}].

\bibitem{BoschiFilho:2002ta}
Henrique Boschi-Filho and Nelson R.~F. Braga.
\newblock Eur. Phys. J. \textbf{C32} (2004): 529-533
  [\href{http://arxiv.org/abs/hep-th/0209080}{arXiv: hep-th/0209080}].

\bibitem{deTeramond:2005su}
Guy~F. de~Teramond and Stanley~J. Brodsky.
\newblock Phys. Rev. Lett. \textbf{94} (2005): 201601
  [\href{http://arxiv.org/abs/hep-th/0501022}{arXiv: hep-th/0501022}].

\bibitem{Brodsky:2006uqa}
Stanley~J. Brodsky and Guy~F. de~Teramond.
\newblock Phys. Rev. Lett. \textbf{96} (2006): 201601
  [\href{http://arxiv.org/abs/hep-ph/0602252}{arXiv: hep-ph/0602252}].

\bibitem{Erlich:2005qh}
Joshua Erlich, Emanuel Katz, Dam~T. Son, and Mikhail~A. Stephanov.
\newblock Phys. Rev. Lett. \textbf{95} (2005): 261602
  [\href{http://arxiv.org/abs/hep-ph/0501128}{arXiv: hep-ph/0501128}].

\bibitem{DaRold:2005zs}
Leandro Da~Rold and Alex Pomarol.
\newblock Nucl. Phys. \textbf{B721} (2005): 79-97
  [\href{http://arxiv.org/abs/hep-ph/0501218}{arXiv: hep-ph/0501218}].

\bibitem{Hirn:2005nr}
Johannes Hirn and Veronica Sanz.
\newblock JHEP \textbf{12} (2005): 030
  [\href{http://arxiv.org/abs/hep-ph/0507049}{arXiv: hep-ph/0507049}].

\bibitem{Hong:2004sa}
Sungho Hong, Sukjin Yoon, and Matthew~J. Strassler.
\newblock JHEP \textbf{04} (2006): 003
  [\href{http://arxiv.org/abs/hep-th/0409118}{arXiv: hep-th/0409118}].

\bibitem{Grigoryan:2007vg}
Hovhannes~R. Grigoryan and Anatoly~V. Radyushkin.
\newblock Phys. Lett. \textbf{B650} (2007): 421-427
  [\href{http://arxiv.org/abs/hep-ph/0703069}{arXiv: hep-ph/0703069}].

\bibitem{Brodsky:2007hb}
Stanley~J. Brodsky and Guy~F. de~Teramond.
\newblock Phys. Rev. \textbf{D77} (2008): 056007
  [\href{http://arxiv.org/abs/0707.3859}{arXiv: 0707.3859}].

\bibitem{Kwee:2007dd}
Herry~J. Kwee and Richard~F. Lebed.
\newblock JHEP \textbf{01} (2008): 027
  [\href{http://arxiv.org/abs/0708.4054}{arXiv: 0708.4054}].

\bibitem{Grigoryan:2007wn}
H.~R. Grigoryan and A.~V. Radyushkin.
\newblock Phys. Rev. \textbf{D76} (2007): 115007
  [\href{http://arxiv.org/abs/0709.0500}{arXiv: 0709.0500}].

\bibitem{Forkel:2007ru}
Hilmar Forkel.
\newblock Phys. Rev. \textbf{D78} (2008): 025001
  [\href{http://arxiv.org/abs/0711.1179}{arXiv: 0711.1179}].

\bibitem{Abidin:2009aj}

\newblock Zainul Abidin and Carl~E. Carlson (2009):
  [\href{http://arxiv.org/abs/0908.2452}{arXiv: 0908.2452}].

\bibitem{Sakai2005a}
Tadakatsu Sakai and Shigeki Sugimoto.
\newblock Prog. Theor. Phys. \textbf{113} (2005): 843-882
  [\href{http://arxiv.org/abs/hep-th/0412141}{arXiv: hep-th/0412141}].

\bibitem{Panico:2007qd}
Giuliano Panico and Andrea Wulzer.
\newblock JHEP \textbf{05} (2007): 060
  [\href{http://arxiv.org/abs/hep-th/0703287}{arXiv: hep-th/0703287}].

\bibitem{Domokos:2007kt}
Sophia~K. Domokos and Jeffrey~A. Harvey.
\newblock Phys. Rev. Lett. \textbf{99} (2007): 141602
  [\href{http://arxiv.org/abs/0704.1604}{arXiv: 0704.1604}].

\bibitem{Pomarol:2008aa}
Alex Pomarol and Andrea Wulzer.
\newblock Nucl. Phys. \textbf{B809} (2009): 347-361
  [\href{http://arxiv.org/abs/0807.0316}{arXiv: 0807.0316}].

\bibitem{Grigoryan:2008up}
H.~R. Grigoryan and A.~V. Radyushkin.
\newblock Phys. Rev. \textbf{D77} (2008): 115024
  [\href{http://arxiv.org/abs/0803.1143}{arXiv: 0803.1143}].

\bibitem{Chernyak:1983ej}
V.~L. Chernyak and A.~R. Zhitnitsky.
\newblock Phys. Rept. \textbf{112} (1984): 173.

\bibitem{Braun:1994ij}
Vladimir~M. Braun and Igor~E. Halperin.
\newblock Phys. Lett. \textbf{B328} (1994): 457-465
  [\href{http://arxiv.org/abs/hep-ph/9402270}{arXiv: hep-ph/9402270}].

\bibitem{Khodjamirian:1997tk}
Alexander Khodjamirian.
\newblock Eur. Phys. J. \textbf{C6} (1999): 477-484
  [\href{http://arxiv.org/abs/hep-ph/9712451}{arXiv: hep-ph/9712451}].

\bibitem{Gokalp:2001sr}
A.~Gokalp and O.~Yilmaz.
\newblock Eur. Phys. J. \textbf{C24} (2002): 117-120
  [\href{http://arxiv.org/abs/nucl-th/0103033}{arXiv: nucl-th/0103033}].

\bibitem{Zhu:1998bm}
Shi-lin Zhu, W.~Y.~P. Hwang, and Ze-sen Yang.
\newblock Phys. Lett. \textbf{B420} (1998): 8-12
  [\href{http://arxiv.org/abs/nucl-th/9802043}{arXiv: nucl-th/9802043}].

\bibitem{Strassler2004}
Sungho Hong, Sukjin Yoon, and Matthew~J. Strassler.
\newblock JHEP \textbf{04} (2006): 003
  [\href{http://arxiv.org/abs/hep-th/0409118}{arXiv: hep-th/0409118}].

\bibitem{Radyushkin2007}
H.~R. Grigoryan and A.~V. Radyushkin.
\newblock Phys. Rev. \textbf{D76} (2007): 095007
  [\href{http://arxiv.org/abs/0706.1543}{arXiv: 0706.1543}].

\bibitem{Ball:1997rj}
Patricia Ball and Vladimir~M. Braun.
\newblock Phys. Rev. \textbf{D55} (1997): 5561-5576
  [\href{http://arxiv.org/abs/hep-ph/9701238}{arXiv: hep-ph/9701238}].

\bibitem{Braun:1988qv}
Vladimir~M. Braun and I.~E. Filyanov.
\newblock Z. Phys. \textbf{C44} (1989): 157.

\bibitem{Eletsky:1982im}
V.~L. Eletsky and Ya.~I. Kogan.
\newblock Z. Phys. \textbf{C20} (1983): 357.

\bibitem{Titov:1999eu}
A.~I. Titov, T.~S.~H. Lee, H.~Toki, and O.~Streltsova.
\newblock Phys. Rev. \textbf{C60} (1999): 035205.

\bibitem{Oh:2000pr}

\newblock Yong-seok Oh, Alexander~I. Titov, and T.~S.~Harry Lee (2000):
  [\href{http://arxiv.org/abs/nucl-th/0004055}{arXiv: nucl-th/0004055}].

\bibitem{Amsler:2008zzb}
C.~Amsler et~al. (Particle Data Group Collaboration).
\newblock Phys. Lett. \textbf{B667} (2008): 1.

\bibitem{Grigoryan:2008cc}
Hovhannes~R. Grigoryan and Anatoly~V. Radyushkin.
\newblock Phys. Rev. \textbf{D78} (2008): 115008
  [\href{http://arxiv.org/abs/0808.1243}{arXiv: 0808.1243}].

\bibitem{Domokos:2009cq}

\newblock S.~K. Domokos, H.~R. Grigoryan, and J.~A. Harvey (2009):
  [\href{http://arxiv.org/abs/0905.1949}{arXiv: 0905.1949}].

\bibitem{Becciolini:2009fu}

\newblock Diego Becciolini, Michele Redi, and Andrea Wulzer (2009):
  [\href{http://arxiv.org/abs/0906.4562}{arXiv: 0906.4562}].

\end{thebibliography}
\end{document}